# A One-Saddle $1/5$ Approximation Algorithm for Common-Kernel Bimatrix Games

Recognition, Exact Segment Optimization, Sharp Selector Bounds, and Certified Robustness

Davit Gondauri

Business and Technology University (BTU), Tbilisi, Georgia

ORCID: 0000-0002-9611-3688 Correspondence: dgondauri@gmail.com

**Abstract**

We develop an algorithmic and certification theory for a fixed-normalization class of symmetric bimatrix games with payoffs

$$R_T = \frac{J + 2T - T^{\mathsf{T}}}{3}, \qquad C_T = R_T^{\mathsf{T}},$$

where $T \in [0,1]^{m\times m}$. The geometric class is defined over the real cube, while all computational inputs and exact certificates use its rational subclass. One auxiliary zero-sum saddle problem for $D_T = 2T - T^{\mathsf{T}}$ yields two symmetric profiles with ordinary additive regret bounds $v/3$ and $(1-v)/2$. Their crossing gives a polynomial-time, fixed-normalization 1/5-approximate Nash equilibrium. We prove tightness for this endpoint-selection policy, give an exact polynomial-time optimization over the segment joining the selected saddle strategies, and make explicit the dependence of that post-processing step on the chosen optimal saddle pair. The class admits unique kernel recovery, exact recognition in $O(m^2)$ rational coordinate operations, affine-cube geometry, and a positive-affine representation of every normalized symmetric game; the latter divides all regrets by three, so the 1/5 guarantee is not a scale-invariant guarantee for arbitrary symmetric games. Exact symmetric-equilibrium computation nevertheless remains PPAD-hard in the class.

For arbitrary square rational games, a nearest-class projection in entrywise maximum norm is a rational linear program with an explicit dual certificate. Combining the projection with the saddle solver gives a certified $(1/5 + 2\eta^*)$-approximate equilibrium, and an a posteriori variant remains valid for any feasible approximate kernel and any exactly evaluated candidate profile. Separately, a full-subset selector game satisfies the sharp regret-to-uniformity inequality $\mathrm{TV}(p,u) \le (3 - 1/q)\rho$, with a matching construction. Conditional fine-grained consequences are isolated from the unconditional theory and are stated only under an explicit, parameter-controlled source-hardness premise. Exact rational finite-instance checks and floating-point HiGHS experiments are reported separately, with the latter treated only as numerical consistency evidence.

# 1 Introduction

An approximation algorithm is most useful when its guarantee is tied to a verifiable structural property of the input game. For bimatrix games, this calls for more than a numerical equilibrium candidate: one would like an exact recognition procedure for the intended class, an explicit strategy construction, and a certificate that remains meaningful when the observed game is only approximately structured. This paper develops those components for a fixed-normalization family of symmetric games generated by a common matrix kernel.

For $T \in [0,1]^{m\times m}$, define the real geometric class

$$\mathcal{C}_{\mathrm{CK}} = \left\{ \left( \frac{J + 2T - T^{\mathsf{T}}}{3}, \frac{J + 2T^{\mathsf{T}} - T}{3} \right) : T \in [0,1]^{m\times m},\ m \geq 1 \right\}. \tag{1}$$

Its rational computational subclass is denoted by $\mathcal{C}_{\mathrm{CK}}^{\mathbb{Q}}$, obtained by requiring $T \in \mathbb{Q}^{m\times m}$. All complexity and exact-algorithm statements below take explicitly encoded rational inputs; the real extension $\mathcal{C}_{\mathrm{CK}}$ is used for affine geometry and compactness.

The main structural question is whether this representation permits efficient additive approximation even though exact symmetric-equilibrium computation remains hard. The answer is affirmative: one auxiliary zero-sum game produces a rational symmetric 1/5-approximate Nash equilibrium in polynomial time.

**Why this class?** The common-kernel representation has an intrinsic payoff geometry that is independent of its historical origin in reduction design. Writing

$$M = \frac{T + T^{\mathsf{T}}}{2}, \qquad S = \frac{T - T^{\mathsf{T}}}{2}, \tag{2}$$

one obtains

$$2T - T^{\mathsf{T}} = M + 3S, \qquad R_T = \frac{J + M + 3S}{3}, \qquad C_T = \frac{J + M - 3S}{3}. \tag{3}$$

Thus the players share the same symmetric baseline $(J + M)/3$, while the antisymmetric component enters with opposite signs. The auxiliary zero-sum direction is therefore not an unrelated algebraic device: it is the amplified oppositional component already present in the payoff geometry. The factor three is exactly the coefficient that links this directional matrix to symmetric regret.

The algorithm is accompanied by a complete input-to-certificate workflow. Membership can be decided exactly and a member has a unique recovered kernel. For games outside the class, an entrywise $\ell_\infty$ projection computes the nearest common-kernel representative and a matching dual witness certifies the projection distance. Exact optimization along the line segment joining one optimal row strategy and one optimal column strategy of the auxiliary zero-sum game can further improve the returned profile.

**Contributions and scope.** The central result is the class-specific one-saddle 1/5 theorem and its constructive proof. The other unconditional results characterize the domain, sharpen the output, or make the guarantee certifiable. A separate selector theorem calibrates a full-subset gadget. The fine-grained compiler consequences are conditional and are separated from the main theory.

| Result | Nature | Assumption | Output |
|---|---|---|---|
| Recognition and recovery | unconditional | rational square game | unique kernel or rejection |
| One-saddle solver | unconditional | $G \in \mathcal{C}_{\mathrm{CK}}^{\mathbb{Q}}$ | fixed-scale 1/5-ANE |
| Segment optimization | unconditional | selected optimal saddle pair | exact best profile on its segment |
| Projection and duality | unconditional | rational square game | distance, kernel, dual certificate |
| Selector calibration | unconditional | full subset selector | sharp coefficient $3 - 1/q$ |
| Compiler exclusion | conditional | explicit source-hardness premise | local reduction-design obstruction |

**Normalization is essential.** The class is geometrically restrictive at the declared $[0, 1]$ scale, but strategically broad under positive affine equivalence: every normalized symmetric game has a positive-affine representative in $\mathcal{C}_{\mathrm{CK}}$. That embedding divides all unilateral gains by three. Consequently, the 1/5 theorem is a genuine statement about the declared common-kernel scale, not a general 1/5 algorithm for arbitrary symmetric games in their original payoff scale.

**Research origin.** The class arose while examining candidate constructions for a deterministic 1/3 approximation-threshold program. The discovery that this region is efficiently approximable turns that candidate architecture into a positive structured-game algorithm and, only secondarily, a conditional diagnostic for reduction design. The unconditional mathematics does not depend on the hardness premise.

The remainder of the paper is organized as follows. Section 2 positions the contribution. Section 3 fixes definitions and the exact computational model. Sections 4 and 5 contain the common-kernel geometry, exact-recognition theory, the 1/5 algorithm, segment post-processing, and robustness certificates. Section 6 gives the selector calibration. Section 7 isolates the conditional fine-grained implications. Section 8 reports exact and numerical verification. Section 9 discusses interpretation, and Section 10 states limitations and open problems. Compiler-specific structural diagnostics are moved to Appendix A.

# 2 Related literature and the specific research gap

## 2.1 Approximation algorithms and zero-sum starting points

Lipton, Markakis, and Mehta established small-support approximation methods [1]. Tsaknakis and Spirakis developed an optimization approach to bimatrix approximation [13]; Deligkas, Fasoulakis, and Markakis later obtained a polynomial-time $(1/3 + \delta)$ guarantee for every fixed $\delta > 0$ [2]. These guarantees are stated for normalized general bimatrix games and, like all additive-regret constants, must be interpreted on their declared payoff scale.

Bosse, Byrka, and Markakis use an auxiliary zero-sum game as a starting point for approximate-equilibrium computation [14]. Thus neither zero-sum initialization nor LP-based approximation is new here. The present result concerns the specific directional matrix $D_T = 2T - T^{\mathsf{T}}$ under the common-kernel constraint and the complementary inequalities $v/3$ and $(1 - v)/2$ that yield the 1/5 policy.

For symmetric games, Kontogiannis and Spirakis use a quadratic formulation and approximate KKT points to obtain $(1/3 + \delta)$-approximate Nash equilibria [4]. Czumaj, Fasoulakis, and Jurdziński study the stronger well-supported notion and obtain a $(1/2 + \delta)$-WSNE for symmetric bimatrix games [5]. The guarantee in this paper is an ordinary additive $\varepsilon$-ANE, not a WSNE.

Automated analysis is another related direction. Deng, Li, and Li study constraint-based approximation analysis [19], while Li, Li, and Deng combine symbolic analysis with language-model-guided discovery in LegoNE [22]. The present paper instead gives explicit class-specific proofs and a prescribed solver; computational experiments are verification aids, not substitutes for the universal argument.

## 2.2 Decomposition, strategic equivalence, recognition, and nearby games

Hwang and Rey-Bellet study decompositions into potential, zero-sum, and related components [9, 10]; Candogan, Menache, Ozdaglar, and Parrilo give a flow-based decomposition into potential, harmonic, and nonstrategic components [15]. The common-kernel parameterization uses the same broad idea of separating cooperative and oppositional structure, but imposes the particular affine cube $0 \leq T \leq 1$ and the inverse $T = 2R + R^{\mathsf{T}} - J$.

Positive affine transformations preserve best-response sets and Nash equilibria; their special role among equilibrium-preserving transformations is discussed explicitly by Tewolde and Conitzer [23]. This matters here because the common-kernel class is a strict fixed-scale subset, while Proposition 4.3 below gives a positive-affine representative of every normalized symmetric game and tracks the exact scaling of additive regret.

Nearest-structured-game analysis also has precedents. Candogan, Ozdaglar, and Parrilo formulate near-potential games and connect payoff-space distance to equilibrium behavior [16]. Our projection specializes the idea to $\mathcal{C}_{\text{CK}}$, uses entrywise maximum distance, and couples the projection to an explicit dual witness and a constructive $1/5 + 2\eta^*$ bound. Neither LP duality nor the generic $2\eta$ perturbation inequality is claimed as new in isolation.

The common-kernel class has no dimension-independent rank bound for $R_T + C_T$, so fixed-rank methods [6] impose a different restriction. Mehta proves, among other results, that computing a symmetric Nash equilibrium of a symmetric two-player game with rank at least six is PPAD-hard [7]. Together with the affine embedding below, this yields exact-equilibrium hardness inside $\mathcal{C}_{\text{CK}}$ despite the constant-error approximation algorithm.

Kontogiannis and Spirakis also analyze strategically-zero-sum bimatrix games [20]. The use of an auxiliary zero-sum game here does not imply that the original common-kernel game is strategically zero-sum. Common-kernel games must likewise be distinguished from symmetric common-payoff games. Ghosh and Hollender show CLS-completeness for computing a symmetric equilibrium in symmetric common-payoff games [21]; within $\mathcal{C}_{\text{CK}}$, common payoffs occur exactly when $T = T^{\mathsf{T}}$.

## 2.3 Selectors, hardness, and the novelty boundary

Subset-selection constructions associated with Althöfer and later hardness reductions control departures from uniform mixing [8, 17]. The selector theorem in Section 6 gives the exact optimal coefficient for one declared payoff family with a neutral action and the full subset space. It does not introduce selector gadgets generally and does not transfer the coefficient to compressed families without a separate proof.

Rubinstein's inapproximability results [17, 18] and Golowich's fine-grained analysis [3] motivate the accounting used in the conditional section, but they do not by themselves instantiate the abstract source premise adopted here. The conditional corollaries therefore remain explicitly parameterized consequences rather than claims that a specific named conjecture has already been mapped into this framework.

**Precise novelty statement.** To the author's knowledge, the literature above does not provide the combination of (i) the explicit common-kernel inverse and recognition test, (ii) the fixed-normalization one-saddle 1/5 certificate, (iii) exact polynomial-time segment optimization for a selected saddle pair, and (iv) the nearest-class primal-dual certificate developed here. This statement concerns that combined construction, not priority for symmetric-game approximation, auxiliary zero-sum methods, game decomposition, positive affine equivalence, or projection-based robustness in general.

| Prior line | aux. zero-sum | exact inverse | 1/5 CK bound | exact segment | dual projection cert. |
|---|---|---|---|---|---|
| Bosse et al. [14] | ✓ | – | – | – | – |
| Game decompositions [10, 15] | related | different | – | – | related |
| Near-potential games [16] | – | different | – | – | related |
| This paper | ✓ | ✓ | ✓ | ✓ | ✓ |

## 3 Definitions and computational framework

The study is deductive and algorithmic. Inputs are explicitly encoded finite rational payoff matrices rather than an empirical dataset. Outputs are rational mixed strategies, structural witnesses, and regret certificates. All polynomial-time claims refer to the complete binary input length, including dimensions, numerators, and denominators.

Let $G = (R, C)$ be a finite rational bimatrix game. Unless otherwise stated, payoffs lie in $[0, 1]$. For mixed strategies $x$ and $y$, define

$$r_R^G(x, y) = \max_i e_i^\top Ry - x^\top Ry, \tag{4}$$

$$r_C^G(x, y) = \max_j x^\top Ce_j - x^\top Cy, \tag{5}$$

and

$$\rho_G(x, y) = \max\{r_R^G(x, y), r_C^G(x, y)\}. \tag{6}$$

A profile is an $\varepsilon$-approximate Nash equilibrium ($\varepsilon$-ANE) if $\rho_G(x, y) \le \varepsilon$. This is the ordinary additive notion, not the stronger well-supported condition.

For a finite action set $A$, a kernel $T : A \times A \to \mathbb{R}$, and distributions $x, y \in \Delta(A)$, write

$$T(x, y) = \sum_{a,b} x_a T(a, b) y_b, \quad T(h, y) = \sum_b T(h, b) y_b, \quad T(x, h) = \sum_a x_a T(a, h). \tag{7}$$

Throughout, $\|A\|_\infty = \max_{i,j} |A_{ij}|$ is the entrywise maximum norm, not the induced row-sum norm, and $\langle A, B\rangle = \sum_{i,j} A_{ij} B_{ij}$. For $p, u \in \Delta([q])$, $\mathrm{TV}(p, u) = \frac{1}{2}\|p - u\|_1$. The matrix $J$ denotes all ones, and $\mathrm{Unif}(S)$ is the uniform distribution on a nonempty finite set $S$. For a rational object $Z$, $\mathrm{bits}(Z)$ denotes its complete reduced binary encoding length.

**Exact computational model.** Recognition and the displayed algebraic identities require $O(m^2)$ rational coordinate operations. Exact zero-sum solving, the projection programs, and the segment minimization are invoked in the standard rational LP/arithmetic model: optimal basic solutions can be chosen rational with encoding length polynomial in the input length [12]. The theoretical certificate layer therefore refers to exact rational feasibility and equality checks. Although the geometric class $\mathcal{C}_{\mathrm{CK}}$ is defined over real kernels, every algorithmic input in this paper is rational and every exact LP certificate is taken in the rational subclass $\mathcal{C}_{\mathrm{CK}}^{\mathbb{Q}}$. Floating-point HiGHS computations in Section 8 are intentionally treated only as numerical consistency checks and are never used as formal optimality certificates.

# 4 Common-kernel geometry, recognition, and exact complexity

For $T \in [0,1]^{m\times m}$ define

$$R_T = \frac{J + 2T - T^{\mathsf{T}}}{3}, \qquad C_T = R_T^{\mathsf{T}} = \frac{J + 2T^{\mathsf{T}} - T}{3}. \tag{8}$$

All entries lie in $[0,1]$. We refer to these as common-kernel factor-three games.

## 4.1 Affine-cube geometry and the factor-three decomposition

For fixed $m$, let $\mathcal{C}_{\mathrm{CK},m}$ denote the real $m \times m$ part of the class; its rational points form the computational subclass $\mathcal{C}_{\mathrm{CK},m}^{\mathbb{Q}}$.

**Proposition 4.1** (Affine-cube geometry)**.** *The map*

$$F_m : [0,1]^{m^2} \to \mathcal{C}_{\mathrm{CK},m}, \qquad T \mapsto (R_T, C_T), \tag{9}$$

*is injective and affine, with affine inverse*

$$T = 2R + R^{\mathsf{T}} - J. \tag{10}$$

*Consequently, $\mathcal{C}_{\mathrm{CK},m}$ is affinely isomorphic to $[0,1]^{m^2}$ and is a compact convex polytope of affine dimension $m^2$ inside the $2m^2$-dimensional ambient bimatrix-payoff space. Within the affine space of symmetric bimatrix games satisfying $C = R^{\mathsf{T}}$, it is full-dimensional. Equivalently,*

$$C = R^{\mathsf{T}}, \qquad 1 \le 2R_{ij} + R_{ji} \le 2 \quad \forall i, j. \tag{11}$$

*Its extreme points are exactly the images of binary kernels $T \in \{0,1\}^{m\times m}$.*

*Proof.* Affineness is immediate. If $R_T = R_{T'}$, then $2R_T + R_T^{\mathsf{T}} - J = T = T'$, proving injectivity and the inverse formula. The image of a compact convex polytope under an injective affine map is an affinely isomorphic compact convex polytope of the same affine dimension, and affine isomorphisms preserve extreme points. The inequalities are precisely the coordinate conditions $0 \le T_{ij} \le 1$ after applying (10). □

Two identities expose the shared and opposed components directly:

$$R_T + C_T = \frac{2(J+M)}{3}, \qquad R_T - C_T = 2S, \tag{12}$$

where $M = (T + T^{\mathsf{T}})/2$ and $S = (T - T^{\mathsf{T}})/2$. Moreover,

$$D_T := 2T - T^{\mathsf{T}} = M + 3S. \tag{13}$$

Thus the factor three is structural: it is the amplification of the antisymmetric component in the directional matrix, while the outer division by three in (8) keeps payoffs in $[0,1]$.

### 4.2 Recognition and unique kernel recovery

**Proposition 4.2** (Recognition and unique kernel recovery)**.** *Let $G = (R, C)$ be an arbitrary rational bimatrix game. Then $G \in \mathcal{C}_{\mathrm{CK}}$ if and only if (equivalently, since $G$ is rational, $G \in \mathcal{C}_{\mathrm{CK}}^{\mathbb{Q}}$ if and only if):*

*(i) $R$ and $C$ are square matrices of the same dimension $m$;*

*(ii) $C = R^{\mathsf{T}}$;*

*(iii) $T_R := 2R + R^{\mathsf{T}} - J$ is entrywise in $[0, 1]$, equivalently $1 \le 2R_{ij} + R_{ji} \le 2$ for all $i, j$.*

*When these conditions hold, $T_R$ is the unique representing kernel. Membership and recovery require $O(m^2)$ rational coordinate operations and polynomial bit complexity in the total input encoding length.*

*Proof.* If $R = (J + 2T - T^{\mathsf{T}})/3$, then $2R + R^{\mathsf{T}} - J = T$, so the kernel is forced. Conversely, if (i)–(iii) hold, substituting $T_R$ into (8) recovers $R$, while (ii) gives $C$. Each entry is obtained by a constant number of rational additions and multiplications by small integers. □

**Algorithm 1: Recognize-and-Recover.** Given rational $(R, C)$: (1) check equal square dimensions and $C = R^{\mathsf{T}}$; (2) compute $T = 2R + R^{\mathsf{T}} - J$; (3) check $0 \le T_{ij} \le 1$ entrywise; (4) return either rejection or the unique rational kernel $T$ together with the equalities and interval tests as a membership certificate.

At the fixed $[0, 1]$ payoff scale, $\mathcal{C}_{\mathrm{CK}}$ is a strict subclass of normalized symmetric games; the zero-payoff symmetric game is an immediate counterexample.

### 4.3 Positive affine embedding and strategic universality

**Proposition 4.3** (Positive affine embedding)**.** *Let $G = (A, A^{\mathsf{T}})$ be a normalized symmetric bimatrix game with $A \in [0, 1]^{m \times m}$. Define*

$$T_A = \frac{2A + A^{\mathsf{T}}}{3}. \tag{14}$$

*Then $T_A \in [0, 1]^{m \times m}$ and its common-kernel game is*

$$\widehat{G} = \left( \frac{J + A}{3}, \frac{J + A^{\mathsf{T}}}{3} \right). \tag{15}$$

*The games $G$ and $\widehat{G}$ have exactly the same best-response correspondences, Nash equilibria, and symmetric Nash equilibria. If $A$ is rational, then $T_A$ and $\widehat{G}$ are rational, so $\widehat{G} \in \mathcal{C}_{\mathrm{CK}}^{\mathbb{Q}}$. For every profile,*

$$\rho_{\widehat{G}}(x, y) = \frac{1}{3} \rho_G(x, y). \tag{16}$$

*Proof.* The matrix $T_A$ is an entrywise convex combination of $A$ and $A^{\mathsf{T}}$, so it lies in $[0, 1]^{m \times m}$. Also $2T_A - T_A^{\mathsf{T}} = A$, which gives (15). Each player's utility undergoes the positive affine transformation $u \mapsto (1 + u)/3$, preserving best-response sets and exact equilibria while dividing every unilateral gain by three. □

**Corollary 4.4** (Strategic universality versus metric restriction)**.** *At the declared normalization,* $\mathcal{C}_{\mathrm{CK}}$ *is a strict geometric subclass of symmetric games. Under positive affine equivalence, however, it contains a representative of every normalized symmetric game. Exact equilibrium structure is preserved by this representation, whereas additive approximation constants are rescaled by a factor of three. The* embedded copy (15) lies *entrywise in the slice* $[1/3, 2/3]$.

*Remark* 4.5 (Normalization dependence). A 1/5-ANE of $\widehat{G}$ becomes only a 3/5-ANE of the original $G$ after inverse rescaling. Thus 1/5 is not a strategic-equivalence invariant constant.

### 4.4 Structural separation and exact-equilibrium hardness

**Proposition 4.6** (Fixed-normalization structural relations)**.** *For* $G_T = (R_T, C_T) \in \mathcal{C}_{\mathrm{CK}}$*:*

*(a)* $R_T = C_T$ *if and only if* $T = T^\top$*; then* $R_T = C_T = (J + T)/3$.

*(b)* $\mathcal{C}_{\mathrm{CK}}$ *has no dimension-independent bound on* $\mathrm{rank}(R_T + C_T)$*. For* $T = I_m$*,* $R_T = C_T = (J + I_m)/3$ *and* $\mathrm{rank}(R_T + C_T) = m$*.*

*(c) Fixed-rank symmetric games are not contained in* $\mathcal{C}_{\mathrm{CK}}$ *at the declared normalization; the zero game is a rank-zero counterexample.*

*Proof.* Part (a) follows from $2T - T^\top = 2T^\top - T$ if and only if $T = T^\top$. For (b), $J + I_m$ has eigenvalues $m + 1$ once and 1 with multiplicity $m - 1$. Part (c) follows from Proposition 4.2. □

**Corollary 4.7** (Exact symmetric-equilibrium hardness survives in the class)**.** *Computing a symmetric Nash equilibrium of an explicitly encoded rational common-kernel factor-three game in* $\mathcal{C}^{\mathbb{Q}}_{\mathrm{CK}}$ *is PPAD-hard.*

*Proof.* Mehta proves PPAD-hardness of computing a symmetric Nash equilibrium in symmetric two-player games of rank at least six [7]. Consider a rational hard instance $(A, A^\top)$. Because rank at least six excludes a constant payoff matrix, its rational range is positive. Subtract the minimum payoff and divide by the rational range, applying the same positive affine transformation to both players; this normalizes $A$ to $[0, 1]$ with polynomial bit complexity and preserves the equilibrium set. Proposition 4.3 then maps the normalized game into $\mathcal{C}_{\mathrm{CK}}$ in polynomial time, again preserving all exact equilibria. A symmetric equilibrium of the image therefore yields one of the original instance. □

## 5 Approximation, exact post-processing, and certified robustness

Let $D_T = 2T - T^\top$ and, for $z \in \Delta([m])$, define

$$\Phi_T(z) = \max_h e_h^\top D_T z. \tag{17}$$

Let

$$v = \min_{z \in \Delta([m])} \Phi_T(z) = \max_{w \in \Delta([m])} \min_{z \in \Delta([m])} w^\top D_T z \tag{18}$$

by minimax. Let $y$ be a minimizing column strategy and $x$ a row maximin strategy for the zero-sum matrix $D_T$.

### 5.1 The symmetric regret identity

**Lemma 5.1** (Symmetric regret identity). *For every $z \in \Delta([m])$, both players have the same regret at $(z,z)$ and*

$$3\rho_{G_T}(z,z) = \Phi_T(z) - T(z,z). \tag{19}$$

*Proof.* Against $z$, the row player's pure-action payoff in $G_T$ is $(1 + e_h^\mathsf{T} D_T z)/3$, whereas the current payoff is $(1 + z^\mathsf{T} D_T z)/3 = (1 + T(z,z))/3$. Subtracting gives (19); symmetry gives the same expression for the column player. □

### 5.2 Complementary saddle branches and the $1/5$ theorem

**Lemma 5.2** (Minimizer branch). *The zero-sum value satisfies $v \geq 0$, and the minimizing strategy $y$ obeys*

$$\rho_{G_T}(y,y) \leq \frac{v}{3}. \tag{20}$$

*Proof.* For every $z$, $\Phi_T(z) \geq z^\mathsf{T} D_T z = T(z,z) \geq 0$, so $v \geq 0$. Since $\Phi_T(y) = v$ and $T(y,y) \geq 0$, Lemma 5.1 gives the claim. □

**Lemma 5.3** (Maximin branch). *The zero-sum value satisfies $v \leq 1$, and the row maximin strategy $x$ obeys*

$$\rho_{G_T}(x,x) \leq \frac{1-v}{2}. \tag{21}$$

*Proof.* For every $w$, $\min_z w^\mathsf{T} D_T z \leq w^\mathsf{T} D_T w = T(w,w) \leq 1$, so $v \leq 1$. Maximin optimality gives $x^\mathsf{T} D_T e_h \geq v$ for every pure $h$. Write $a = T(h,x)$ and $b = T(x,h)$. Then $2b - a \geq v$, hence $b \geq (a+v)/2$, and therefore

$$e_h^\mathsf{T} D_T x = 2a - b \leq \frac{3a - v}{2} \leq \frac{3 - v}{2}. \tag{22}$$

Taking the maximum gives $\Phi_T(x) \leq (3-v)/2$. Averaging $x^\mathsf{T} D_T e_h \geq v$ with weights $x_h$ yields $T(x,x) = x^\mathsf{T} D_T x \geq v$. Lemma 5.1 now gives

$$3\rho_{G_T}(x,x) \leq \frac{3-v}{2} - v = \frac{3(1-v)}{2}. \tag{23}$$

□

**Theorem 5.4** (One-saddle factor-three algorithm). *For every rational $T \in [0,1]^{m \times m}$, the value $v$ lies in $[0,1]$ and*

$$\rho_{G_T}(y,y) \leq \frac{v}{3}, \qquad \rho_{G_T}(x,x) \leq \frac{1-v}{2}. \tag{24}$$

*Hence the threshold policy*

$$\mathcal{A}(T) = \begin{cases} (y,y), & v \leq 3/5, \\ (x,x), & v > 3/5 \end{cases} \tag{25}$$

*returns a rational symmetric profile satisfying*

$$\rho_{G_T}(\mathcal{A}(T)) \leq \frac{1}{5}. \tag{26}$$

*The algorithm runs in time polynomial in the complete rational encoding length of the game, equivalently of its recovered kernel.*

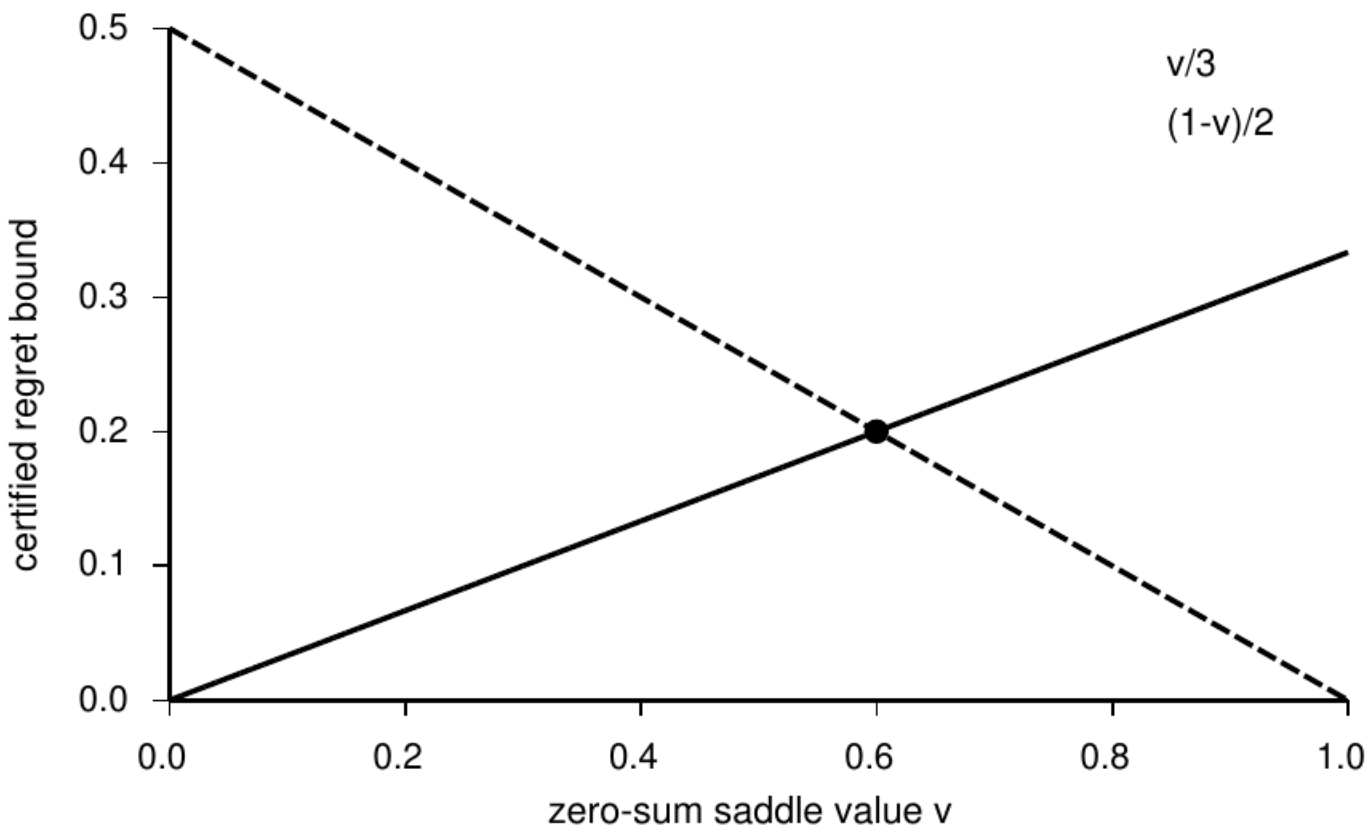

Figure 1: The two certified branch bounds. Their unique crossing occurs at $v = 3/5$ and height $1/5$.

*Proof.* Lemmas 5.2 and 5.3 give the two bounds. Their upper envelopes cross when $v/3 = (1 - v)/2$, i.e. at $v = 3/5$, where both equal $1/5$. Exact rational zero-sum linear programming returns rational optimal strategies and value with polynomial encoding length [12]. □

**Algorithm 2: Common-Kernel Saddle Solver.** Given $T$ (or a game accepted by Algorithm 1): (1) form $D_T = 2T - T^{\mathsf{T}}$; (2) solve the single auxiliary zero-sum game, obtaining a row maximin strategy $x$, a column minimizer $y$, and value $v$ (primal and dual LPs may be solved separately); (3) return $y$ if $v \le 3/5$ and $x$ otherwise; (4) optionally apply the exact segment optimizer in Theorem 5.7.

*Remark* 5.5 (Scope of the constant). Theorem 5.4 is a fixed-normalization guarantee. It does not prove that $1/5$ is the optimal approximation constant attainable in polynomial time on $\mathcal{C}_{\mathrm{CK}}$.

### 5.3 Tightness of the endpoint-selection analysis

**Proposition 5.6** (Policy-level tightness)**.** *There are two families of $2 \times 2$ kernels indexed by $\varepsilon \in (0, 2/5)$ whose selected-branch regrets equal $1/5 - \varepsilon/3$ and $1/5 - \varepsilon/2$, respectively. Therefore*

$$\sup_T \rho_{G_T}(\mathcal{A}(T)) = \frac{1}{5} \tag{27}$$

*for the endpoint policy* (25)*, with the supremum approached already by $2 \times 2$ kernels. This is a policy-level tightness result, not a computational lower bound for all algorithms on $\mathcal{C}_{\mathrm{CK}}$.*

*Proof.* Let $v^- = 3/5 - \varepsilon$ and

$$T_\varepsilon^- = \begin{pmatrix} 0 & 1 \\ (1 + v^-)/2 & v^- + \varepsilon/2 \end{pmatrix}. \tag{28}$$

Then

$$D_\varepsilon^- = \begin{pmatrix} 0 & (3 - v^-)/2 \\ v^- & v^- + \varepsilon/2 \end{pmatrix}. \tag{29}$$

For a column mixture $(1 - t, t)$, the second row receives $v^- + t\varepsilon/2$, so every $t > 0$ has maximum row payoff strictly above $v^-$. Thus $y = e_1$ is the unique minimizer, the value is $v^-$, and $T_{11} = 0$. Lemma 5.1 gives

$$\rho_{G_{T_\varepsilon^-}}(y, y) = \frac{v^-}{3} = \frac{1}{5} - \frac{\varepsilon}{3}. \tag{30}$$

For $v^+ = 3/5 + \varepsilon < 1$, let

$$T_\varepsilon^+ = \begin{pmatrix} 0 & 1 \\ (1+v^+)/2 & v^+ \end{pmatrix}. \tag{31}$$

The strategy $e_2$ guarantees $v^+$ against both columns. A row mixture placing positive mass $t$ on $e_1$ receives $(1-t)v^+ < v^+$ against column $e_1$, hence $x = e_2$ is the unique row maximin strategy. Since $v^+ < 1$, the selected regret is

$$\rho_{G_{T_\varepsilon^+}}(x, x) = \frac{1 - v^+}{2} = \frac{1}{5} - \frac{\varepsilon}{2}. \tag{32}$$

Letting $\varepsilon \downarrow 0$ proves the claim. □

### 5.4 Exact segment-optimized post-processing

The endpoint policy uses only $x$ or $y$, but the same saddle computation also supplies the segment connecting them.

**Theorem 5.7** (Exact segment optimization for a selected saddle pair)**.** *Let $T$ be rational and let $x$ and $y$ be rational row-maximin and column-minimizing strategies returned by an exact saddle computation for $D_T$. Define*

$$z_\lambda = \lambda x + (1-\lambda)y, \qquad \lambda \in [0,1]. \tag{33}$$

*There is a deterministic polynomial-time algorithm that computes a rational $\lambda^* \in [0,1]$ satisfying*

$$\rho_{G_T}(z_{\lambda^*}, z_{\lambda^*}) = \min_{\lambda \in [0,1]} \rho_{G_T}(z_\lambda, z_\lambda). \tag{34}$$

*Moreover,*

$$\rho_{G_T}(z_{\lambda^*}, z_{\lambda^*}) \le \min\{\rho_{G_T}(x,x), \rho_{G_T}(y,y)\} \le \frac{1}{5}. \tag{35}$$

*Proof.* By Lemma 5.1,

$$3\rho_{G_T}(z_\lambda, z_\lambda) = \max_h e_h^\top D_T z_\lambda - T(z_\lambda, z_\lambda). \tag{36}$$

For each pure action $h$, the function $\lambda \mapsto e_h^\top D_T z_\lambda$ is affine with rational coefficients. Its maximum over $h$ is a continuous piecewise-affine upper envelope. Candidate breakpoints occur among the $O(m^2)$ pairwise intersections of these affine functions, and every such rational intersection has polynomial encoding length. The function $\lambda \mapsto T(z_\lambda, z_\lambda)$ is a rational quadratic polynomial. Between consecutive candidate breakpoints, the ordering of all affine functions is fixed, so the active maximum is a single affine function and the regret is one rational quadratic polynomial.

On each interval, the minimum is attained at an endpoint or at the stationary point of a nonconstant quadratic when that point lies in the interval; constant and linear pieces are handled by endpoints. Parallel lines create no isolated intersection and identical lines may be deduplicated. Thus polynomially many rational candidates suffice, and exact comparison identifies a global minimizer with polynomial bit complexity. Finally, $\lambda = 0$ and $\lambda = 1$ recover $y$ and $x$. □

**Algorithm 3: Segment-Optimize.** Construct the $m$ affine functions $a_h(\lambda) = e_h^\top D_T z_\lambda$. Enumerate their pairwise intersections in $[0,1]$, add 0 and 1, sort the rational candidates, and identify the active affine branch on each interval. Subtract the common quadratic $T(z_\lambda, z_\lambda)$, add every in-interval stationary point, and exactly evaluate all candidates. Return the minimizing rational $\lambda^*$.

*Remark* 5.8 (Dependence on the selected optimal saddle pair). Theorem 5.7 is conditional on the particular optimal pair $(x, y)$ returned by the zero-sum solver. When the saddle point is nonunique, different choices $x \in X^*$ and $y \in Y^*$ may generate different segments and different optimized regrets. The theorem makes no claim that an arbitrary returned pair is globally optimal over all saddle pairs.

**Example 5.9** (Strict segment improvement)**.** Let

$$T = \begin{pmatrix} 3/4 & 3/4 \\ 1 & 1/4 \end{pmatrix}. \tag{37}$$

The zero-sum matrix $D_T$ has value $v = 1/2$, with row maximin $x = e_1$ and column minimizer $y = e_2$. The endpoint regrets are $1/6$ and $1/12$. At $\lambda = 1/3$,

$$z_{1/3} = (1/3, 2/3) \tag{38}$$

is an exact symmetric Nash equilibrium: both pure row payoffs equal $19/36$, and symmetry gives the same for the column player. Thus segment post-processing can reduce positive endpoint regret all the way to zero.

## 5.5 Perturbation stability

**Lemma 5.10** (Entrywise payoff perturbation)**.** *Let $G = (R, C)$ and $G' = (R', C')$ have the same action sets and satisfy $\|R' - R\|_\infty \le \delta$ and $\|C' - C\|_\infty \le \delta$. Every $\varepsilon$-ANE of $G$ is an $(\varepsilon + 2\delta)$-ANE of $G'$.*

*Proof.* For either player, every pure-deviation payoff changes by at most $\delta$, so the best-response value changes by at most $\delta$. The payoff of the current mixed profile also changes by at most $\delta$. Hence each regret increases by at most $2\delta$. □

## 5.6 Nearest-class projection and explicit dual certificate

For a rational square game $G' = (R', C')$, define the projection distance

$$\begin{aligned} \eta^*(G') = \min_{T,\eta} \quad & \eta \qquad (39) \\ \text{s.t.} \quad & 0 \le T_{ij} \le 1 \quad \forall i, j, \\ & -\eta \le R'_{ij} - \frac{1 + 2T_{ij} - T_{ji}}{3} \le \eta \quad \forall i, j, \\ & -\eta \le C'_{ij} - \frac{1 + 2T_{ji} - T_{ij}}{3} \le \eta \quad \forall i, j, \\ & \eta \ge 0. \qquad (40) \end{aligned}$$

**Proposition 5.11** (Algorithmic projection)**.** *Program* (40) *has polynomial size and computes*

$$\eta^*(G') = \min_{G_T \in \mathcal{C}_{\text{CK}}} \max\{\|R' - R_T\|_\infty, \|C' - C_T\|_\infty\}. \tag{41}$$

*It returns an attaining rational kernel $T^*$ in polynomial time, with polynomial output bit length.*

*Proof.* Each absolute-value condition is represented by two rational linear inequalities. There are $m^2 + 1$ variables and $O(m^2)$ constraints. The feasible region is nonempty (for example, choose any $T \in [0, 1]^{m \times m}$ and sufficiently large $\eta$), and standard rational LP theory supplies an optimal rational solution of polynomial bit length [12]. □

Let

$$B_R = R' - \frac{J}{3}, \qquad B_C = C' - \frac{J}{3}, \tag{42}$$

and define

$$Q_{ij}(U,V) = \frac{2U_{ij} - U_{ji} - V_{ij} + 2V_{ji}}{3}. \tag{43}$$

**Proposition 5.12** (Explicit dual certificate)**.** *The projection distance admits the dual representation*

$$\eta^*(G') = \max_{\|U\|_1+\|V\|_1\le 1} \left\{ \langle U, B_R\rangle + \langle V, B_C\rangle - \sum_{i,j}[Q_{ij}(U,V)]_+ \right\}, \tag{44}$$

*where $[a]_+ = \max\{a,0\}$ and $\|U\|_1 = \sum_{i,j}|U_{ij}|$. Equivalently, with $U = U^+ - U^-$ and $V = V^+ - V^-$, the dual is the rational LP*

$$\begin{aligned} \max \quad & \langle U^+ - U^-, B_R\rangle + \langle V^+ - V^-, B_C\rangle - \sum_{i,j} W_{ij} && (45)\\ s.t. \quad & \sum_{i,j}(U^+_{ij} + U^-_{ij} + V^+_{ij} + V^-_{ij}) \le 1, \\ & W_{ij} \ge Q_{ij}(U^+ - U^-, V^+ - V^-) \quad \forall i,j, \\ & U^+, U^-, V^+, V^-, W \ge 0. && (46) \end{aligned}$$

*A primal feasible pair $(T^*, \eta^*)$ and a dual feasible tuple with exactly equal rational objective values form a directly checkable certificate of the exact nearest-class distance. Optimal primal and dual certificates may be chosen rational with polynomial encoding length.*

*Proof.* Vectorize $T$ and the two residual matrices. Program (40) is

$$\min_{0\le T\le 1} \|b - \mathcal{A}T\|_\infty, \quad b = (B_R, B_C), \quad \mathcal{A}T = \left(\frac{2T - T^\top}{3}, \frac{2T^\top - T}{3}\right). \tag{47}$$

Using $\|w\|_\infty = \max_{\|u\|_1\le 1}\langle u, w\rangle$ gives a convex-concave bilinear optimization over the compact convex kernel cube and compact convex $\ell_1$ unit ball. The objective is continuous and affine in each variable, so finite-dimensional minimax interchange applies. Hence

$$\eta^*(G') = \max_{\|u\|_1\le 1}\left(\langle u, b\rangle - \max_{0\le T\le 1}\langle u, \mathcal{A}T\rangle\right). \tag{48}$$

Writing $u = (U,V)$, the coefficient of $T_{ij}$ in $\langle u, \mathcal{A}T\rangle$ is exactly $Q_{ij}(U,V)$. Therefore

$$\max_{0\le T\le 1}\langle u, \mathcal{A}T\rangle = \sum_{i,j}[Q_{ij}(U,V)]_+, \tag{49}$$

which proves (44). Splitting signed variables and linearizing the positive part with $W$ yields (46). Strong LP duality and rational bit-complexity bounds complete the certificate statement [12]. □

**Algorithm 4: Project-and-Solve.** For rational square $G'$: (1) solve (40) exactly to obtain $(T^*, \eta^*)$; (2) solve (46) exactly and verify rational equality of the primal and dual objectives; (3) solve $G_{T^*}$ by Theorem 5.4, optionally using Theorem 5.7; (4) return the symmetric profile, the kernel witness, and the projection certificate.

**Theorem 5.13** (Polynomial-time neighborhood algorithm)**.** *For every rational square game $G'$, Algorithm 4 returns a rational profile that is a*

$$\left(\frac{1}{5} + 2\eta^*(G')\right)\text{-}ANE \tag{50}$$

*of $G'$ in time polynomial in the complete rational input encoding length.*

*Proof.* Theorem 5.4 gives a $1/5$-ANE of $G_{T^*}$. Both payoff matrices of $G'$ are entrywise within $\eta^*(G')$ of the projected game. Apply Lemma 5.10. □

**Proposition 5.14** (A posteriori certificate without projection optimality)**.** *Let $\widetilde{T} \in [0,1]^{m\times m}$ be any feasible rational kernel and set*

$$\widehat{\eta} = \max\{\|R' - R_{\widetilde{T}}\|_\infty, \|C' - C_{\widetilde{T}}\|_\infty\}. \tag{51}$$

*For any rational symmetric profile $(z,z)$,*

$$\rho_{G'}(z,z) \le \rho_{G_{\widetilde{T}}}(z,z) + 2\widehat{\eta}. \tag{52}$$

*Thus neither an optimal projection nor an exact saddle solution is required to certify a particular candidate once $\widetilde{T}$, $z$, and the two regrets on the right-hand side are evaluated exactly. Optimality of $\widehat{\eta}$ requires the separate dual witness of Proposition 5.12.*

*Proof.* Apply Lemma 5.10 to $G_{\widetilde{T}}$ and $G'$ with $\delta = \widehat{\eta}$. □

*Remark* 5.15 (Meaning of the 1/15 scale)*.* For normalized input games, the projection certificate falls strictly below the numerical 1/3 level whenever

$$\frac{1}{5} + 2\eta^* < \frac{1}{3}, \qquad \text{equivalently} \qquad \eta^* < \frac{1}{15}. \tag{53}$$

The value 1/15 is therefore the natural certified radius generated by the present 1/5 core and the generic $2\eta$ perturbation transfer. It is not asserted to be a sharp geometric boundary.

# 6 Sharp full-subset selector calibration

Fix $q \ge 2$. The row player chooses $i \in [q]$. The column player chooses either a neutral action $\perp$ or an arbitrary subset $S \subseteq [q]$. Define

$$A(i,\perp) = \frac{1}{2}, \qquad A(i,S) = 1 - \mathbf{1}[i \in S], \tag{54}$$

and

$$B(i,\perp) = \frac{1}{2}, \qquad B(i,S) = \frac{1}{2} + \frac{1}{2}\left(\mathbf{1}[i \in S] - \frac{|S|}{q}\right). \tag{55}$$

These payoffs lie in $[0,1]$.

Let $p$ be the row distribution and $u = (1/q,\dots,1/q)$. The standard total-variation characterization gives

$$d := \mathrm{TV}(p,u) = \max_{S\subseteq[q]} \left\{p(S) - \frac{|S|}{q}\right\}. \tag{56}$$

Indeed, for any signed vector $p - u$ with zero total mass, the maximizing set is $S = \{i : p_i > 1/q\}$.

For a mixed column strategy $\sigma$, define $a_i = \Pr_{S\sim\sigma}[i \in S]$, with the neutral action contributing zero; let $m_0 = \min_i a_i$ and $b_i = a_i - m_0 \ge 0$.

**Lemma 6.1** (Exact selector regrets)**.** *For every mixed profile,*

$$r_R = p^\mathsf{T} b, \qquad 2r_C = d - c, \qquad c = (p-u)^\mathsf{T} b. \tag{57}$$

*Proof.* If the neutral action has probability $\lambda$, pure row $i$ receives $1-\lambda/2-a_i$, so a best row chooses an index attaining $m_0$ and the row regret is $p^\mathsf{T}(a - m_0\mathbf{1}) = p^\mathsf{T} b$. A pure subset $S$ gives the column player payoff

$$\frac{1}{2} + \frac{1}{2}\left(p(S) - \frac{|S|}{q}\right), \tag{58}$$

so the best pure payoff is $1/2 + d/2$. The expected excess above $1/2$ under $\sigma$ is

$$\frac{1}{2}\sum_i \left(p_i - \frac{1}{q}\right) a_i = \frac{1}{2}(p-u)^\mathsf{T} b, \tag{59}$$

because $(p-u)^\mathsf{T}\mathbf{1} = 0$. Hence $2r_C = d - c$. □

Since $p_i \le 1$ and $b_i \ge 0$,

$$p^\mathsf{T} b \le \sum_i b_i = q\, u^\mathsf{T} b, \tag{60}$$

and therefore

$$c = p^\mathsf{T} b - u^\mathsf{T} b \le \left(1 - \frac{1}{q}\right) p^\mathsf{T} b. \tag{61}$$

**Theorem 6.2** (Sharp finite-$q$ selector inequality)**.** *For every mixed profile,*

$$\mathrm{TV}(p,u) \le \left(1 - \frac{1}{q}\right) r_R + 2r_C \le \left(3 - \frac{1}{q}\right)\rho_G. \tag{62}$$

*The optimal universal constant $C_q$ in $\mathrm{TV}(p,u) \le C_q\rho_G$ for this declared payoff family is exactly*

$$C_q = 3 - \frac{1}{q}. \tag{63}$$

*Proof.* From the lemma, $d = c + 2r_C$. Substitute the bound on $c$ and then use $r_R, r_C \le \rho_G$. For sharpness, set $p = e_1$ so $d = (q-1)/q$, and let the column player use $S = \{1\}$ with probability $t$ and $\perp$ otherwise. Then

$$r_R = t, \qquad r_C = \frac{q-1}{2q}(1-t). \tag{64}$$

Balancing the regrets gives $t = (q-1)/(3q-1)$ and

$$\frac{d}{\rho_G} = \frac{3q-1}{q} = 3 - \frac{1}{q}. \tag{65}$$

□

*Remark* 6.3 (Representation boundary). The theorem uses all $2^q$ subsets, so the column has $2^q + 1$ actions. It is an exact calibration result rather than a size-efficient selector when $q$ grows. Restricted or compressed Althöfer-style constructions require their own constants [8, 17].

# 7 Conditional implications for reduction design

This section is logically separate from Sections 4–6. Nothing in the recognition, approximation, segment, projection, or selector theorems depends on the following premise.

## 7.1 Parameter-controlled fine-grained source model

**Assumption 7.1** (FG: uniformly encoded source hardness). *Let $\mathcal{H}$ be a uniformly encoded source family. For $H \in \mathcal{H}$, let $\mathrm{enc}(H)$ be its binary encoding and let $n = n(H)$ be the native hardness parameter of the intended source theorem. Assume polynomial equivalence between parameter and input length: there exist fixed constants $c_1, c_2 > 0$ such that, for all sufficiently large instances,*

$$n \le |\,\mathrm{enc}(H)|^{c_1}, \qquad |\,\mathrm{enc}(H)| \le n^{c_2}. \tag{66}$$

*Let $\beta(n)$ have polynomial-size encoding, and let $\mathrm{Sol}_\beta(H)$ be a total source-solution relation whose valid outputs have encoding length $n^{O(1)}$. Assume no deterministic algorithm solves every instance in time $2^{o(n)}$.*

*For each fixed $K$, an admissible compiler is one uniform deterministic algorithm that outputs an explicitly encoded rational bimatrix game $G_K(H)$ in time $2^{o(n)}$ and with complete binary length*

$$N_K(H) = \mathrm{bits}(G_K(H)) = 2^{o(n)}. \tag{67}$$

*A total decoder $D_K$ is defined on every rational mixed profile and runs in time polynomial in $|\,\mathrm{enc}(H)| + N_K(H) + \mathrm{bits}(x, y)$.* (68)

**Lemma 7.2** (Subexponential accounting). *Under Assumption 7.1, if $N_K(H) = 2^{o(n)}$ and a rational LP or arithmetic routine runs in time polynomial in $N_K(H) + |\,\mathrm{enc}(H)|$ and returns an output of polynomial bit length in those quantities, then the routine and the decoder together run in $2^{o(n)}$ time.*

*Proof.* For fixed constants $a, b$, $(2^{o(n)})^a n^b = 2^{o(n)}$. Equation (66) replaces every polynomial factor in the explicit source encoding by $2^{o(n)}$. Standard rational LP output-size bounds keep $\mathrm{bits}(x, y)$ polynomial in the explicit compiled input length, so decoder time remains $2^{o(n)}$. □

*Remark* 7.3 (Scope). Assumption 7.1 is deliberately abstract and does not claim that ETH for PPAD, Rubinstein's constant-error result, or Golowich's vanishing-error framework automatically gives the exact premise above. Any concrete application must identify the source family, source accuracy, native parameter, and mapping required by the specific hardness theorem [17, 3].

## 7.2 Threshold bridge and exact class exclusion

Fix $K \ge 8$ and set

$$\alpha_K = \frac{1}{3} - \frac{1}{K}. \tag{69}$$

**Definition 7.4** (Threshold bridge). An admissible compiler-decoder pair $(G_K, D_K)$ satisfies the threshold bridge at $\alpha_K$ if, for every source instance $H$ and every rational mixed profile $(x, y)$ of the compiled game,

$$\rho_{G_K(H)}(x, y) \le \alpha_K \quad \Longrightarrow \quad D_K(H, x, y) \in \mathrm{Sol}_\beta(H). \tag{70}$$

**Corollary 7.5** (Exact common-kernel compiler exclusion). *Under Assumption 7.1, no admissible compiler whose output lies in $\mathcal{C}_{\mathrm{CK}}^{\mathbb{Q}}$ for every source instance can satisfy the threshold bridge at $\alpha_K$ for any fixed $K \ge 8$.*

*Proof.* Recognize and recover $T$ by Proposition 4.2, then compute a rational 1/5-ANE by Theorem 5.4. Since $1/5 < 1/3 - 1/K$ for $K \ge 8$, the bridge decoder returns a source solution. Compilation, exact LP solving, and decoding together take $2^{o(n)}$ time by Lemma 7.2, contradicting Assumption 7.1. □

**Corollary 7.6** (Certified neighborhood exclusion)**.** *Under Assumption 7.1, an admissible square-game threshold-bridge compiler at $\alpha_K$ cannot satisfy, for every source instance,*

$$\eta^*(G_K(H)) \le \delta_K, \qquad 0 \le \delta_K \le \frac{1}{15} - \frac{1}{2K}, \tag{71}$$

*for any fixed $K \ge 8$.*

*Proof.* Theorem 5.13 returns a rational profile with regret

$$\frac{1}{5} + 2\eta^*(G_K(H)) \le \frac{1}{5} + 2\delta_K \le \frac{1}{3} - \frac{1}{K} = \alpha_K. \tag{72}$$

The threshold bridge decodes a source solution, and Lemma 7.2 again gives an overall $2^{o(n)}$ source algorithm. □

The radius $1/15 - 1/(2K)$ is sufficient, not claimed maximal. It equals $1/240$ at $K = 8$ and tends to $1/15$ as $K \to \infty$.

# 8 Computational verification and reproducibility

The theoretical results above rest on exact rational proofs. The companion release **CKNASH-2026.09.27-r3** separates exact finite-instance verification from floating-point consistency checks and supplies all source files, machine-readable outputs, a Git source bundle, licenses, a reproduction wrapper, and a single SHA-256 manifest. The main script is `common_kernel_reproducibility.py`; its output is `verification_results.json`. The random seed is 20260927 and the numerical tolerance is $10^{-8}$.

The exact layer checks the selector equality cases $q$ in {2,3,5,10,50}, the worked segment example, and the complete binary kernel families used for finite auditing. For $m = 4$, the dedicated C++ verifier enumerates all $2^{16} = 65{,}536$ binary kernels, constructs an exact saddle certificate by integer determinant/adjugate arithmetic, verifies the global saddle inequalities, checks 0<=v<=1 and both branch inequalities, and evaluates the selected 1/5 rule exactly. All 65,536 instances pass with zero missing saddle certificates and zero branch or selected-bound failures.

A separate exact-rational Segment-Optimize verifier covers all 528 binary kernels in dimensions $m = 2,3$. Across those instances it processes **1,098 breakpoint entries in aggregate**, 570 open intervals, 94 accepted stationary candidates, and 1,124 candidate points. Assertions verify breakpoint equalities, active-envelope consistency, stationary-point interval membership, exact endpoint dominance, and nonnegative regret; all pass. The largest segment-optimized regret in this finite set is 1/12, and 480 tested segments attain zero regret.

The numerical layer uses NumPy and SciPy/HiGHS only as a consistency check. It tests 200 denominator-20 random kernels, 14 threshold-family instances approaching v = 3/5, and 30 perturbed-game projection instances. Floating-point LP output is never treated as an exact primal-dual certificate; exact certification remains the rational procedure of Proposition 5.12.

| Test | Cases | Arithmetic | Observed criterion |
|---|---|---|---|
| Selector sharpness | 5 values of q | exact rational | exact equality d/rho = 3 - 1/q |
| Segment worked example | 1 | exact rational | regret 0; both pure payoffs 19/36 |
| Binary saddle check | 528 kernels (m < = 3) | exact rational | all branch and 1/5 bounds pass |
| Full binary m = 4 saddle / 1/5 audit | 65,536 kernels | exact integer/ rational | 0 missing certificates; 0 branch failures; 0 selected-bound failures; largest selected regret 1/6 |
| Segment internal assertions | 528 kernels (m < = 3) | exact rational | 1,098 aggregate breakpoint entries; 570 intervals; 94 stationary candidates; 1,124 candidates; all assertions pass |
| Random saddle check | 200 | floating HiGHS | max branch violation 0 at printed precision; max selected regret 1/6; 25 strict 101-grid segment improvements |
| Threshold families | 14 | floating HiGHS | max selected regret 0.1999999967; below 1/5 |
| Projection | 30 | floating HiGHS | max primal-dual gap 1.11e-16; max feasibility residual 1.67e-16; no transfer-bound violation |

The exhaustive figures above are finite verification statistics, not strengthened worst-case theorems. In particular, the universal 1/5 result remains Theorem 5.4, and the exact segment guarantee remains Theorem 5.7.

**Environment and integrity.** The reference numerical run used Python 3.13.5, NumPy 2.3.5, SciPy 1.17.0 and HiGHS 1.8.0 on Linux x86-64. The dedicated exact m = 4 verifier is C++17; the segment verifier uses Python `Fraction` arithmetic and writes to a relative output path (or to a path supplied with `--output`), so no host-specific absolute output path is required. The complete package is checked by `verify_manifest.py` against `SHA256SUMS.txt`.

# 9 Interpretation and possible uses

## 9.1 Input-to-certificate workflow

For square rational payoffs on a declared scale, first apply exact recognition. If the game is in $\mathcal{C}_{\mathrm{CK}}$, recover $T$, solve one auxiliary zero-sum game, and optionally optimize the selected saddle segment. If the game is outside the class, compute a nearest common-kernel representative and its dual distance certificate. Finally, evaluate the returned profile directly in the original game. The resulting record contains a strategy, a structural witness, a payoff-space discrepancy, and a certified regret bound.

For example, a certified projection distance $\eta^* = 0.01$ gives the deterministic bound $0.2 + 0.02 = 0.22$. Direct evaluation may be better, and Proposition 5.14 allows the actually computed regret to replace the worst-case $1/5$ term whenever it is known exactly.

## 9.2 What projection distance does and does not mean

The quantity $\eta^*$ measures entrywise payoff-space closeness to $\mathcal{C}_{\mathrm{CK}}$. It does not by itself imply behavioral closeness of equilibria, economic similarity of models, or closeness of equilibrium correspondences. Those are separate claims requiring additional regularity or application-specific analysis.

If estimated payoffs are within $\tau$ entrywise of the true payoff matrices and the estimated game is within $\eta^*$ of $\mathcal{C}_{\mathrm{CK}}$, Lemma 5.10 gives the deterministic bound

$$\frac{1}{5} + 2\eta^* + 2\tau. \tag{73}$$

Here $\eta^*$ measures structural misspecification relative to the common-kernel class, whereas $\tau$ measures estimation uncertainty. A statistical interpretation of $\tau$ requires its own sampling model and confidence argument; this paper does not supply one.

## 9.3 Stylized finite-action interpretation

A two-action instance may be read as two agents choosing from the same policy menu. Example 5.9, for instance, may be relabeled "policy 1" and "policy 2" without altering any mathematics: exact recognition identifies the kernel, the auxiliary saddle produces endpoint candidates, and the segment optimizer identifies an exact mixed equilibrium. This is a worked mathematical interpretation, not an empirical calibration. Applications to markets, networks, or multi-agent systems require a defensible payoff model and a declared normalization.

# 10 Limitations and theorem-driven open problems

The results should be read as a structured-game approximation and certification theory. The principal limitations are the following.

First, the $1/5$ guarantee is tied to the fixed common-kernel normalization and is not invariant under arbitrary positive affine rescaling. Second, Proposition 5.6 proves tightness only for the endpoint policy, not for all polynomial-time algorithms on $\mathcal{C}_{\mathrm{CK}}$. Third, Theorem 5.7 optimizes only the segment generated by a selected optimal saddle pair; nonuniqueness of the saddle strategies leaves a larger optimization problem. Fourth, the transfer $1/5 + 2\eta^*$ uses the generic $2\eta$ perturbation inequality and may be non-sharp. Fifth, the projection is defined in the entrywise $\ell_\infty$ norm; other norms induce different geometric and algorithmic questions. Sixth, exact primal-dual certification is a rational-arithmetic statement, while the supplied HiGHS runs are numerical consistency checks. Seventh, the selector result uses the exponential full-subset action family. Eighth, no statistical payoff-estimation model or empirical dataset is part of the theorem.

The immediate research agenda can be stated as concrete mathematical problems.

**Open Problem 10.1** (Worst-case guarantee for a specified saddle selector)**.** Fix a deterministic polynomial-time rule $S$ that returns one optimal saddle pair $(x_S(T), y_S(T)) \in X^*(T) \times Y^*(T)$ for each rational kernel $T$. Define

$$z_{S,\lambda}(T) = \lambda x_S(T) + (1-\lambda) y_S(T). \tag{74}$$

Determine

$$\sup_{m\geq 1} \sup_{T\in[0,1]^{m\times m}\cap\mathbb{Q}^{m\times m}} \min_{\lambda\in[0,1]} \rho_{G_T}(z_{S,\lambda}(T), z_{S,\lambda}(T)), \tag{75}$$

and decide whether there exists such a polynomial-time selector $S$ whose uniform guarantee is strictly below $1/5$. This makes saddle-pair selection part of the algorithm rather than leaving the segment ambiguous when the zero-sum optimum is nonunique.

**Open Problem 10.2** (Optimizing over the saddle polytopes)**.** Let $X^*$ and $Y^*$ be the row and column optimal-strategy polytopes of $D_T$. Determine whether

$$\min_{x\in X^*,\ y\in Y^*,\ \lambda\in[0,1]} \rho_{G_T}(\lambda x + (1-\lambda)y, \lambda x + (1-\lambda)y) \tag{76}$$

can be computed in polynomial time and whether optimizing the pair improves the uniform worst-case constant.

**Open Problem 10.3** (Optimal approximation constant on $\mathcal{C}_{\mathrm{CK}}$)**.** Determine the best worst-case additive approximation guarantee attainable in polynomial time on the fixed-normalization class $\mathcal{C}_{\mathrm{CK}}$. The present paper establishes an upper bound of $1/5$ but no matching lower bound.

**Open Problem 10.4** (Sharper robustness radius)**.** Can the coefficient 2 in Lemma 5.10 be reduced using the special geometry of the nearest common-kernel projection, thereby enlarging the certified $\eta^* < 1/15$ neighborhood?

**Open Problem 10.5** (Compressed selector)**.** Can a polynomial-size selector family enforce a regret-to-uniformity inequality with a coefficient comparable to the sharp full-subset value $3 - 1/q$?

The conditional compiler diagnostics in Appendix A raise additional construction questions, but they are not needed to interpret or validate the positive common-kernel algorithm.

# 11 Conclusion

The paper establishes a direct constructive theorem for a declared family of symmetric bimatrix games: exact recognition and kernel recovery reduce the structured input to one auxiliary zero-sum game, whose saddle value yields complementary regret bounds and a fixed-normalization $1/5$-approximate Nash equilibrium. The constant comes from the crossing of the two branch inequalities, and exact optimization along the segment of a selected saddle pair can only improve the returned profile.

The algorithm extends beyond exact membership through a nearest-class linear program with an explicit dual witness. This gives a fully checkable rational certificate of structural distance and, together with entrywise perturbation stability, a $(1/5 + 2\eta^*)$ approximation guarantee for arbitrary square rational games. The full-subset selector theorem supplies a separate sharp calibration result. Exact symmetric-equilibrium computation remains PPAD-hard in the class, clarifying the distinction between exact hardness and fixed-error tractability.

The most immediate unresolved question is no longer how to obtain the endpoint $1/5$ certificate, but how much of that constant can be improved by exploiting the full saddle polytopes rather than one selected pair and two endpoints. Determining the worst-case guarantee of that stronger post-processing problem is the natural next step.

# A Reduction-theoretic structural diagnostics

The results in this appendix arose from the broader threshold-reduction program. They are retained because they provide concise structural warnings for reduction design, but they are not part of the proof of the main approximation theorem.

### A.1 Exact max-regret normal form

With $M = (T + T^{\mathsf{T}})/2$ and $S = (T - T^{\mathsf{T}})/2$, define

$$\Phi_T(z) = \max_h \{M(h, z) + 3S(h, z)\}. \tag{A.1}$$

For a pair $(x, y)$ set

$$A_T(x, y) = \Phi_T(y) - M(x, y) - 3S(x, y) = 3r_R, \tag{A.2}$$

$$B_T(x, y) = \Phi_T(x) - M(x, y) + 3S(x, y) = 3r_C, \tag{A.3}$$

and let $\Gamma_T = A_T + B_T$ and $\Delta_T = A_T - B_T$. Then

$$6\rho_G(x, y) = \Gamma_T(x, y) + |\Delta_T(x, y)|. \tag{A.4}$$

This identity is diagnostic: a soundness argument about maximum regret cannot replace directional control by a stronger symmetric-sum condition without justification.

### A.2 Diffuse valid-support obstruction

Let $U$ be a finite candidate set and $\varnothing \neq V \subseteq U$. Define

$$T^*(u, v) = \mathbf{1}\{v \notin V \text{ or } u = v\}. \tag{A.5}$$

For mixed $x, y$, put

$$p = x(U \setminus V), \quad q = y(U \setminus V), \quad s = \sum_{v \in V} x_v y_v, \tag{A.6}$$

and $m_y = \max_{v \in V} y_v$, $m_x = \max_{v \in V} x_v$.

**Proposition A.1** (Diffuse valid-support obstruction)**.** *For the ideal kernel above,*

$$3r_R = p + m_y - s, \qquad 3r_C = q + m_x - s. \tag{A.7}$$

*Moreover, for every nonempty $S \subseteq V$, the symmetric profile $x = y = \mathrm{Unif}(S)$ is an exact Nash equilibrium. Hence semantic validity alone cannot force a constant-mass valid atom.*

*Proof.* One has $T^*(x, y) = q + s$ and $T^*(y, x) = p + s$. A valid action $h$ has directional payoff $2q + y_h$, whereas an invalid action has payoff $2q - 1$. Since $V$ is nonempty, the best directional payoff is $2q + m_y$. Subtracting the expected directional payoff $2q - p + s$ gives the row formula; symmetry gives the column formula. For $x = y = \mathrm{Unif}(S)$, $p = q = 0$ and $m_x = m_y = s = 1/|S|$, so both regrets vanish. □

### A.3 Threshold-searchability

Let $A_H : \Delta(A_R(H)) \times \Delta(A_C(H)) \to [0, 1]$ be a semantic statistic and define

$$S_H(a_0) = \{(x, y) : A_H(x, y) \geq a_0\}. \tag{A.8}$$

Call the high-region family uniformly subexponentially searchable if one deterministic algorithm, given $\mathrm{enc}(H)$ and the explicit compiled representation, outputs a rational point in $S_H(a_0)$ whenever the set is nonempty, in $2^{o(n)}$ time and with output length $2^{o(n)}$.

**Observation A.2** (Searchability obstruction)**.** *Under Assumption 7.1, the following cannot all hold for every source instance: (i) $S_H(a_0)$ is nonempty; (ii) the family is uniformly subexponentially searchable; and (iii) every rational point of $S_H(a_0)$ decodes into $\mathrm{Sol}_\beta(H)$.*

*Proof.* Compile $H$, run the uniform high-region finder, and decode the rational output. Lemma 7.2 makes the total running time $2^{o(n)}$, contradicting Assumption 7.1. □

This is a composition diagnostic rather than a construction theorem: a mixed-only or nonconvex-looking statistic is not automatically search-hard; the relevant question is the complexity of finding a point in its superlevel region.

## Code and Data Availability

No external dataset is used. The final archival reproducibility release is **CKNASH-2026.09.27-r3** (27 September 2026). The manuscript does not depend on an external project repository: the submission ZIP itself is the canonical distributed object. Its contents include the main verification script and JSON output, the exhaustive exact $m=4$ C++ verifier and recorded result, the portable exact Segment-Optimize verifier and JSON output, a POSIX-shell reproduction wrapper, a cross-platform SHA-256 manifest checker, local licensing notices, the manuscript PDF, and a clean Git source bundle.

The Git bundle is `CKNASH_reproducibility_2026.09.27-r3.bundle` with HEAD `3f9fd45565b4c1c18caf6d837270ceb2b88ae03b`. The main numerical suite uses seed 20260927 and tolerance 1e-8. Exact finite checks use integer or rational arithmetic; HiGHS results are numerical consistency evidence only.

| Core artifact | SHA-256 |
|---|---|
| `common_kernel_reproducibility.py` | `3b71efea0645730ee4849d59dc40ae30`<br>`df657f981a2d54454f3f656520281bbe` |
| `verification_results.json` | `feadd6d83fe17e7578a4c5cd89643fa0`<br>`5ed4badd5781912b9ccbd1fef4534351` |
| `cknash_exact_m4.cpp` | `745263a5e0943ae9ce973c0c1b072c35`<br>`5ad19a7eaf8b0e91a720a8986c125aa4` |
| `cknash_segment_exact_verify.py` | `831d0b0487522a82eea14fb6bd791030`<br>`74c29d4a76fdfee6aca1d7553b289c4c` |
| `CKNASH_reproducibility_2026.09.27-r3.bundle` | `2705b72d12d4d880ebb7161342d5dac2`<br>`415b53b66ea00b5cac54bf87ab989482` |

The final `SHA256SUMS.txt`, generated only after the canonical PDF is fixed, records the digest of every distributed artifact except the manifest itself. Thus the PDF inside the ZIP and the separately distributed canonical PDF are required to be byte-identical. `verify_manifest.py` performs a platform-independent check of that condition.

**Local reproducibility script.** The supplied `reproduce_cknash.sh` first verifies the distributed manifest and then writes reproduction outputs to separate `reproduced_*` files, using only local package paths:

```
#!/usr/bin/env bash
set -euo pipefail
PYTHON=${PYTHON:-python3}; CXX=${CXX:-g++}
"$PYTHON" verify_manifest.py
"$PYTHON" common_kernel_reproducibility.py --output reproduced_verification_results.json
"$PYTHON" cknash_segment_exact_verify.py --output reproduced_segment_exact_results.json
"$CXX" -O3 -std=c++17 cknash_exact_m4.cpp -o cknash_exact_m4.repro
./cknash_exact_m4.repro > reproduced_cknash_exact_m4_results.txt
rm -f cknash_exact_m4.repro
```